\documentclass{article} 
\usepackage{main,times}

\usepackage{amsmath,amsfonts,bm}

\def\eqref#1{equation~\ref{#1}}

\def\1{\bm{1}}

\DeclareMathAlphabet{\mathsfit}{\encodingdefault}{\sfdefault}{m}{sl}
\SetMathAlphabet{\mathsfit}{bold}{\encodingdefault}{\sfdefault}{bx}{n}

\usepackage{booktabs}
\usepackage{hyperref}
\usepackage{url}
\usepackage{siunitx}
\usepackage{booktabs}
\usepackage[titletoc,title]{appendix}
\usepackage{array}
\usepackage{graphicx}
\newcolumntype{C}[1]{>{\centering\arraybackslash}p{#1}}
\usepackage{enumitem}

\usepackage{caption}
\title{Towards Standardization of Data Licenses: The Montreal Data License}

\iclrfinalcopy
\author{Misha Benjamin$^1$, Paul Gagnon$^1$, Negar Rostamzadeh$^1$, Chris Pal$^{1,2,3}$, Yoshua Bengio$^{3,4,5}$, Alex Shee$^1$\\
$^1$ \textnormal{Element AI}\\
$^2$ \textnormal{Polytechnique Montréal}\\
$^3$ \textnormal{MILA}\\
$^4$ \textnormal{Canada CIFAR AI Chair}\\
$^5$ \textnormal{Senior CIFAR Fellow}}

\iclrfinalcopy 
\begin{document}
\maketitle
\begin{abstract}
This paper provides a taxonomy for the licensing of data in the fields of artificial intelligence and machine learning. The paper's goal is to build towards a common framework for data licensing akin to the licensing of open source software. Increased transparency and resolving conceptual ambiguities in existing licensing language are two noted benefits of the approach proposed in the paper. In parallel, such benefits may help foster fairer and more efficient markets for data through bringing about clearer tools and concepts that better define how data can be used in the fields of AI and ML. The paper's approach is summarized in a new family of data license language - \textit{the Montreal Data License (MDL)}. Alongside this new license, the authors and their collaborators have developed a web-based tool to generate license language espousing the taxonomies articulated in this paper~(\url{www.montrealdatalicense.com}).
\end{abstract}

\section{"Data as Oil" – a Flawed but Instructive Analogy}
This paper provides a taxonomy for the licensing of data in the fields of artificial intelligence and machine learning. The paper's goal is to build towards a common framework for data licensing akin to the licensing of open source software. Increased transparency and resolving conceptual ambiguities in existing licensing language are two noted benefits of the approach proposed in the paper. In parallel, such benefits may help foster fairer and more efficient markets for data through bringing about clearer tools and concepts that better define how data can be used in the fields of AI and ML. The paper's approach is summarized in a new family of data license language - the Montreal Data License (MDL). Alongside this new license, the authors and their collaborators have developed a web-based tool to generate license language espousing the taxonomies articulated in this paper~(\url{www.montrealdatalicense.com})\footnote{The authors welcome feedback on the contents of this article, which can be sent via email to~\url{info@montrealdatalicense.com}}.

\textit{Data is the new oil} is an oft-repeated mantra, used in many fora in recognition of the fundamental role that data plays as a catalyst for the creation of artificial intelligence and machine learning assets and systems\footnote{The authors espouse the commonly held view that "artificial intelligence" or "AI" is a catch-all term that refers to a growing number of techniques and methods, themselves often using approaches such as machine learning ("ML") and deep learning. The terms "AI" and "ML" will be used throughout.}. Oil is an appealing analogy from a conceptual standpoint. Its extraction is a resource-heavy endeavor. The processing and refinement of oil yields fuel, plastics and other economically valuable derivatives. The acquisition and "extraction" processes for data are equally resource-intensive, and, given the technological progress underlying the current age of big data, machine learning and artificial intelligence, it is no exaggeration to compare the economic benefits of data in terms comparable to those used for oil. 

In comparison to data, the market for oil is heavily regulated throughout its supply and extraction chain. Regulation is driven by the need to reduce transaction friction, prevent security issues, regulate toxicity and greenhouse gas emissions, and ultimately, to foster public trust. Through regulation and standardization, the end-products derived from oil gained in quality and predictability~\citep{Economist}. One underestimated benefit arising from such regulation is the emergence and consolidation of standardized terminology, which in turn is ultimately market-making and fosters scalability. Unfortunately, none of those elements exist with regards to markets for data - which creates a lot of friction and uncertainty and increase transaction costs. These transaction costs consist of resources allocated to harmonizing the data itself as well as assessing whether the data itself is usable (both technically and from a legal standpoint). While metadata made available alongside data may prove useful in reducing such costs, metadata is inconsistent and may not always facilitate the assessment of how relevant or technically useful data may be. However, metadata is technical in nature, and does not clarify how data may legally be used. Conversely, recent efforts in standardizing the presentation of metadata for ML models are highly relevant, and contribute to fostering transparency in the fields of ML and AI~\citep{gebru2018datasheets, mitchell2019model}. For ML and AI to continue their growth, and for that growth to beneficial for all, standardized terminology and increased predictability is necessary. This article aims to provide a first step towards such standards with respect to data licensing.

The databases that guided the early development of AI models were assembled by university researchers mostly by scraping the internet, or even through other corpus made available, such as the evidence produced in the infamous Enron case. This was necessary given that no dedicated machine learning databases existed at the time, and there was very little funding available to those researchers. From a legal standpoint, depending on the nature of the data, collating and unifying such data in databases could arguably have qualified (under certain legal systems) as copyright infringement or database right infringement (for jurisdictions such as the European Union where such a right exists). In many jurisdictions, fair use-type exemptions under copyright laws very well may have applied to much of the research done in the early days of AI research, and may still continue to apply to some of the research done today. That being said, relying on those exemptions today as a general rule today may be a flawed approach that creates important downsides. 

Data is valuable for the purposes of ML and AI when it is voluminous, organized and, ideally, labelled and tagged. Even in our supposed age of big data, it is more common to find unstructured, messy and noisy data, especially in commercial contexts focused on applying ML and AI-based tools to real-world issues. The biggest companies in the world suffer least from such limitations – their own scale and data-generating capabilities through massive use of their platforms mean they collect and use data on a scale that most other market participants cannot match. For the benefits of ML and AI to be accessible and benefit a wider realm of humanity, other market participants need to be on a level playing field. One way to bridge this gap is through a more transparent, predictable market for data with clear legal language as its underpinning.

In fundamental research, standardized datasets are more common, and may be used to train, develop and improve different ML and AI based tools. These datasets also become benchmarks in their own right, used to measure the performance and accuracy of different tools. However, the right to use these datasets remains an open question in many cases. More pragmatically, the following questions need be answered: where is this data coming from? Can it be used for a proposed use-case? Are there any resulting obligations? This seems intuitive enough, but as this paper will show, it can be a trying exercise rife with conceptual ambiguities.

This paper aims to illustrate the difficulties caused by the current database terms and suggests a resolution through taxonomy better suited to machine learning and artificial intelligence. 

To do so, this paper will first define how data can be "used" in the context of ML and AI. From this diagnosis, we will identify the different components that make up typical databases and the assets that can result from use of databases when data is used for artificial intelligence. The resulting rights to such assets stemming from this analysis are also proposed. This will allow us to propose a framework that would allow database creators to easily generate license language as well as a summary of that language, which could include restrictions commonly found on database use. 

\section{"Use" in ML and AI}
\subsection{Issues with Existing Database Licenses}
A review of some of the databases most commonly used in fundamental AI research reveals a patchwork of language permitting certain uses of the database as a whole, a few examples of which are shown in Appendix 1.

On a side-by-side level, one of the issues that is quickly apparent is that the permissions given to those databases are vaguely worded, and it is not clear whether many of the use cases that are commonly contemplated by database users fall within those permissions. There is no innate predictability as to what license terms could apply at first glance – further analysis is needed. Thus, this creates friction, or, in economic terms, transaction costs as embodied in the time, resources and skill to appropriately ascertain whether a certain database can be used. The following are illustrative examples of the conceptual ambiguities present in commonly used license terms, which, as discussed, the taxonomy presented in this paper aims to resolve.
\begin{enumerate}[label=(\roman*)]
\item
\textbf{Lack of Nuance on ``Use''}
\newline
Licenses typically grant the right to "use" data - without regard to what "use" actually means. Indeed, licenses may define use-cases or restrictions for "research use" or a "commercial use". However, devoid of further context, the licenses all posit one homogenous notion of "use". This forgoes the intricacies of ML and AI. The taxonomy suggested in this paper aims to surmount this false binary of "use or non-use" to propose a more contextually appropriate taxonomy. 
Additionally, copyright law may provide exemptions based on "fair use" or "fair dealing". The applicability of fair use-type exemptions is dependent on a variety of factors (including the use, the type of underlying data, what elements, if any, are commercialized after use of the database etc.), the evaluation of which has not been tested in front of courts in the context of ML and AI. This means that each user has to guess whether or not the database creator has the rights necessary in order to license them to the user, and make an evaluation whether or not their use would be protected by a fair use-type exemption applies to them, which is not at all settled in law for applications of ML and AI. 
\item
\textbf{Commercial vs Non-Commercial Use}
\newline
By way of example, many of the licenses cited in Appendix 1 contain a restriction against "commercial use". This restriction is, in the opinion of the authors, problematically ambiguous. If contracts are about intent, it is impossible to decipher what is meant by restrictions on "commercial use" within database licenses. 

When assessing commercial use, whose intent or purpose is considered? If an employee participates in fundamental research, but is employed by a for-profit entity, should it be considered commercial? Conversely, consider a researcher within a university's whose technology transfer office may patent or commercialize uses of a dataset – would this otherwise academic context be considered commercial? What about the timing of such determination – could an initially non-commercial use be later requalified? If so, by whom? The database administrator may have incentives to reinterpret license terms, potentially for economic gains. Dynamics such as these were once rife in standard-setting bodies, and gave rise to substantial litigation. 

Another example of how the restriction against commercial use is problematic is the use-case of a for-profit company working for a non-profit on a specific project. Could it use the database to train a model for use by a non-profit? If so, could it reuse that model for other purposes after? What if it generated significant positive publicity or goodwill for the for-profit company? 
\item
\textbf{Research}
\newline
The development of ML and AI faces tremendous constraints on talent to drive the changes, and, more importantly, to ensure that such talent meaningfully participates in transmitting such knowledge through universities and other academic fora. Indeed, there is a world-wide hunt for talent that make even the most prestigious universities uncompetitive in the face of high salaries and access to world beating computing resources. The world-leading ML and AI conferences host researchers from more traditional academic context, but, increasingly, from private research groups of for-profit companies. There are inherent differences in context within these companies, such that "commercial" or "non-commercial" may also be a false dichotomy under which "research" is but one use-case. Moreover, researchers from academia may themselves partake in research in collaboration with peers working within for-profit companies, thus blurring the lines further. 

In this context, some database licenses limit the purpose for which they are licensed to "research" use. Without further clarification, this may also problematic given that research is but the first step towards product and solution development. By restricting use of data to "research", is the intent then to posit "research" in opposition to "commercialization of research products"? Some license language goes further and uses "academic research" – but truly this muddies the waters further depending on one's interpretation of "academic" as being context-dependent (i.e. within academia as traditionally understood) or as a substantive qualification of the research work itself. 
\item
\textbf{Lack of Uniformity}
\newline
Another apparent issue with the terms detailed in Appendix 1 is that there is relatively no uniformity or standard terminology used across these different licenses. 

Free and Open Source Software (FOSS) has come of age since its inception over thirty years ago. FOSS communities have truly built a commons: the non-rivalrous and non-excludable characteristics of public goods are readily espoused by most, if not all, FOSS licenses. While there exist a great number of FOSS licenses that can be used, commonly used licenses have centered around a relatively limited number of FOSS licenses, and further splintering or creation of new licenses is generally discouraged.

This standardization of FOSS licenses is one of the reasons that FOSS has trended towards as frictionless use as is possible – there exists a relatively limited number of licenses, grouped into families, that are commonly used. This allows developers to very quickly have clarity on what they can and cannot do with the software in question. Even when uploaders decide to create sui generis licenses for their software, they generally iterate on commonly used licenses, and there is more clarity around what certain terms mean. Thus, at a glance, the mere mention of the kind of license gives clarity to users of FOSS as to what constraints apply to their use, and transaction costs for FOSS (time to analyze, time to adapt) trends towards a minimum. Consider the prevalence and importance of FOSS – without such standardization, it would have been near-impossible to build and leverage the power of decentralized creation and community. 

Moreover, FOSS licenses do not limit the kind of uses that can be made of the underlying materials – the materials are free to use in whatever way the licensee sees fit. Infrequently, a distinct licensing model for commercial use can be spelled out by licensors. However, this is stated clearly. More importantly, \textbf{commercial use} in the context of FOSS is defined much clearer by licensors that choose a dual-licensing model. Thresholds of use, clearly stated commercial use cases and other clarifications are made. Consider, however, that the term â\textbf{use} for software is conceptually sound when considered in contrast to âuse of data in ML and AI. This means that the difficulties encountered in standardizing FOSS license terms may be heightened for datasets used in ML and AI. 
\item
\textbf{Share-Alike Requirements}
\newline
There are datasets that are licensed under licensing terms that bring functional ambiguities that are difficult to reconcile with uses cases in ML and AI. One such license requirement is the "share alike" requirement that is found under the Creative Commons Share Alike license (CC-SA). The CC-SA, as with most licenses of the Creative Commons family, is particularly well-suited to be used in conjunction with copyrighted works in creative fields. The share alike requirement prevents downstream users from re-appropriating licensed content under other license terms. For example, one cannot repackage a collection of photographs licensed under CC-SA and resell it online, or make it available under restrictive terms. 

For applications of ML and AI based upon datasets, the CC-SA creates issues. For one, if a researcher combines CC-SA datasets with other datasets, the share alike requirement may apply to the combined dataset, given that the combined dataset may be qualified as a â\textbf{derivative work} of the initial license. 

More importantly, the notion of â\textbf{derivative work} is ill defined. Indeed, it is unclear whether the transformative nature of ML and AI create such âderivative works, as commonly understood, meaning that it is highly uncertain that representations, models and/or their weights and hyperparameters would be considered âderivative work subject to the share-alike requirement.  
\item
\textbf{Licensing Language Requires Standardization and Context-Appropriate Adaptations to ML and AI}
\newline
Hence, whatever the concern a database administrator has with respect to how databases are used, conceptually adapted database licenses are necessary. Without adapted licenses that are representative of use cases in AI and ML, uncertainty abounds, and less scrupulous and/or more established actors are favored. If the choice of a license to data aims to reflect concerns about commercialization, either because considerable resources are invested in creating databases and need be compensated, or because there are philosophical objections to commercialization, such considerations merit clarity to be truly effective. Not providing such clarity either under deters investment in valuable dataset creation, or tolerates uses cases that run counter to licensors' intent. 

Furthermore, the notion of \textbf{use}â without more is no friend to ML and AI, and the progress made in ML and AI should also be reflected by licenses that reflect the iterative process that move fundamental research to commercially available products and solutions, akin to the foundational moments of license creation for FOSS. More granularity can better reflect licensors' intent, and in turn, ensure that the progress in ML and AI is not premised upon ill-adapted licenses to the data driving such progress. 
\end{enumerate}

\subsection{Consequences of uncertainty}
The consequences of the uncertainty related to usage rights are manifold. For starters, many actors in the AI community are forced to spend time and effort attempting to interpret vague language and make risk assessments based on that uncertainty. Depending on how the terms are interpreted, the user may (i) refrain from using a database that it would have the right to use, resulting in less productivity, less advancement of research, lower quality of data (which can often cause more bias or unfairness) and/or more data acquisition costs; or (ii) use a data in violation of the terms of the dataset - this can result in privacy-related harms and copyright violations, as well as lower the incentives and affect the market for companies whose business model is to make available datasets while respecting data and privacy rights in the underlying data points. 

This creates perverse incentives, where the more scrupulous actors actually have less data to work with and are less successful, whereas the less scrupulous actors benefit from the uncertainty - effectively making a disregard for copyright and privacy a competitive advantage. In an era when there is a lot effort from regulators, government, tech companies and NGOs devoted to ensuring respect for data, and particularly personally identifiable information (PII), creating a market distortion where violations of those rights are rewarded seems particularly regrettable. 

Another problematic effect of that uncertainty is that it disproportionately affects small and medium enterprises (SMEs). Universities could come to rely on certain fair use-type exemptions and may be less likely to face litigation related to their research activities. Very early stage startups can bet the company to use unclean datasets because they have very little to lose and much to gain, and large incumbent companies can afford to litigate against data subjects, have other sources of data and have patents and legal teams that have a chilling effect against such suits. SMEs on the other hand, have to justify their use of data sets to investors, and can't afford to take the risks that the other actors can. 

\section{Proposed Taxonomy for Standardized AI Rights to Data}
\subsection{Baseline Definitions}
How is data "used" in ML and AI? What does it mean when it is stated that algorithms need be exposed to data in order to fundamentally achieve their purpose and harness the potential of the advanced techniques being discussed? We propose the following definitions:
\begin{itemize}
\item 
\textbf{Data:} By "data", we refer to the information being made available. The format and layout of such information is referred to as a database or dataset, whatever way it may be organized. Where applicable, the data may be separated into different segments into underlying data or metadata in the form of data tags and other structural information. Such data may be collected and harnessed from different sources, or made available from a single-source. Data can be basic collated information (e.g. a range of measurements such as temperature, location) or be formed of more complex information (e.g. pictures, maps).
\item
\textbf{Train:} To expose an Untrained Model to the Data in order to adjust the weights, hyperparameters and/or structure thereof.

\item
\textbf{Labelled Data:} Datasets are especially useful when their metadata is highly organized, namely in the form of labels or tags. This kind of metadata may be created manually, automatically, by users or crowdsourced. The original uploader of the data set may also have labelled or tagged the data, but this can evolve over time or be compounded. Labels are metadata that can allow Models to perform better and ascribe meaning or representations to the underlying data. For example, if a dataset of 10,000 images is tagged with "dog" or "no dog", and a vision-based model is trained to identify the presence of dogs from such dataset, the presence of the labels may prove essential to the exercise, depending on the methodology used to train such model. 
\item
\textbf{Model:} The term "Model" is used to refer to ML or AI-based algorithms, or assemblies thereof that, in combination with different techniques, may be used to obtain certain results. These results can be insights on past data patterns, predictions on future trends or more abstract results. Different learning techniques exist, and are used to accomplish different tasks. The term "Model" is thus inherently generic and is made to capture a wide breadth of techniques, though, admittedly, conceptual and practical differences are immense.
\item
\textbf{Untrained Model:} A model is deemed untrained when such model has been devised and adjustments have been made to its structure and components. These adjustments are made after exposing different variations of the model to the Data in order to find the optimal model, but that model has not been trained on the Data as such. On a technical level, with regards to a given set of Data, the Untrained Model is the version of the model for which the weights and hyperparameters have not been adjusted using that Data. 
\item
\textbf{Trained Model:} An Untrained Model that has been modified by exposure to the Data. In common parlance, a trained model has "learned" from fulsome exposure to data. There may be many training phases resulting in tweaks to the model itself, or to the Data. 
\item
\textbf{Representation:} A different form, format or model that mimics the effects of given data, but that does not contain any individual data points or allow third parties to infer individual data points with currently existing technology. Basically, a Trained Model need not necessarily require constant exposure to Data itself, and a transposed input of Data that does actually not carry it over is referred to as a Representation.
\item
\textbf{Output:} The output is defined as the "results" of applying a Trained Model to data. Thus, the output depends on the Model and what is sought in a particular use case. For example, the output of temperature predicting models would be temperature prediction data, or the output from a retailer's pricing models could be the estimated price that consumers are willing to pay. However, within ML and AI, the outputs may also be far-reaching: automatically generated tags and images, even art, can be a model's output. Output is essentially what is sought from using a Trained Model. 
\end{itemize}

\subsection{New Taxonomy Describing "Use" of Data in AI and ML}
As described above, it is clear that use of data in ML and AI is an incremental process that reaches from extraction and refinement to actionable output. Along such value chain, different actions may be taken with data: it is "used" in different ways, for different purposes. The below cases illustrate how data is "used" within ML and AI. These use cases are the foundation of the Montreal Data License, the proposed text of which is found at Appendix 4. Where the definitions appear prescriptive, or be deliberately mutually exclusive, the reader should understand that this was done to clearly delineate use cases. 
\begin{table}[]
\begin{tabular}{ll}
\multicolumn{2}{c}{\textbf{The Data itself}} \\
\\
\hline
\multicolumn{1}{|p{7cm}|}{\textit{Access}} & \multicolumn{1}{|p{7cm}|}{To access, view and/or download the Data to view it and evaluate it (evaluation algorithms may be exposed to it, but no Untrained Models).} \\ \hline
\multicolumn{1}{|p{7cm}|}{\textit{Labelling}} & \multicolumn{1}{|p{7cm}|}{To build upon Data by adding tags, labels or other metadata to the dataset or subsets of the Data.} \\ \hline
\multicolumn{1}{|p{7cm}|}{\textit{Distribute}} & \multicolumn{1}{|p{7cm}|}{Make all or part of the Data available to third parties.} \\ \hline
\multicolumn{1}{|p{7cm}|}{\textit{Represent}} & \multicolumn{1}{|p{7cm}|}{Transform the data into a new representation, thereby re-representing each data element in a way that mimics the effects of the initial data itself (i.e. the purpose or end-result consists of a suitable alternative to such Data).} \\ \hline \\
\multicolumn{2}{c}{\textbf{Use of the Data in Conjunction with Models}} \\ \\ \hline
\multicolumn{1}{|p{7cm}|}{\textit{Benchmark (case 1: without training a model, Case 2: where a model is trained on the data so as to evaluate it)}} & \multicolumn{1}{|p{7cm}|}{To access the Data, use the Data as training data to evaluate the efficiency of different Untrained Models, algorithms and structures, but excludes reuse of the Trained Model, except to show the results of the Training. This includes the right to use the dataset to measure the performance of a Trained or Untrained Model, without however having the right to carry-over weights, code or architecture or implement any modifications resulting from such evaluation.} \\ \hline
\multicolumn{1}{|p{7cm}|}{\textit{Research}} & \multicolumn{1}{|p{7cm}|}{To access the Data, use the Data to create or improve Models, but without the right to use the Output or resulting Trained Model for any purpose other than evaluating the Model Research under the same terms.} \\ \hline
\multicolumn{1}{|p{7cm}|}{\textit{Publish}} & \multicolumn{1}{|p{7cm}|}{To make available to third parties the Models resulting from Research, provided however that third parties accessing such Trained Models have the right to use them for Research or Publication only.} \\ \hline
\multicolumn{1}{|p{7cm}|}{\textit{Internal Use}} & \multicolumn{1}{|p{7cm}|}{To access the Data, use the Data to create or improve Models and resulting Output, but without the right to Output Commercialization or Model Commercialization. The Output can be used internally for any purpose, but not made available to Third Parties or for their benefit.} \\ \hline
\multicolumn{1}{|p{7cm}|}{\textit{Output Commercialization}} & \multicolumn{1}{|p{7cm}|}{To access the Data, use the Data to create or improve Models and resulting Output, with the right to make the Output available to Third Parties or to use it for their benefit. The Trained Model itself however cannot be not made available to Third Parties. This would allow SaaS commercialization.} \\ \hline
\multicolumn{1}{|p{7cm}|}{\textit{Model Commercialization}} & \multicolumn{1}{|p{7cm}|}{Make a Trained Model itself available to a Third Party, or embodying the Trained Model in a product or service, with or without direct access to the Output for such Third Party.} \\ \hline
\end{tabular}
\end{table}
\newpage
\subsection{Illustration of How the Taxonomy is Applied}
In order to illustrate the above uses, we can take the example of a database of historical equities trades (prices, volumes etc.). 

With the right to Evaluate models, a licensee working on a trading models (the Model) would be able to take any existing (untrained/pretrained) versions of the Model it has and train them on all or part of that data in order to test its performance (i.e. the quality of the Output) on historical data. They could also test a variety of different Models in this way in order to choose which is the best for given circumstances. They could also iteratively make changes to the architecture/code of the Model, and test how those different iterations perform, and would not have any restrictions on using those versions of the of the Model. However, they could not use any of the Output to inform any stock trades, nor reuse the Model as modified by access to the Data (ex. the weights of the Trained Model would have to be deleted).

With the right to use the Data for Research, the licensee would be entitled to evaluate the Model, but would also be entitled to use the Trained Model and/or the Output to create other datasets and trained Models, \textit{provided} that all such Trained Models and datasets are subject to the same restrictions. In short, this allows the licensee to play around with the data and resulting models in order to see what is possible, but not to make any trades or offer any investment advice based on the outcome of that research without separate permission from the licensor. The authors see this as a useful licensing model for data providers who want to allow AI companies to develop products that could consume their data (and therefore create a new market for their data) without an initial contractual commitment on the AI company's part, but the AI Company would then have to negotiate a paid license to actually use the product of that research. 

The right to Publish with regards to a Model would give the licensee to do everything that would be allowed under a Research license, but would, in addition, allow the licensee to publish all the product of the Research, provided that the all such product of Research is available under the same restrictions as those related to the original Data. This is what data licensors often refer to when they refer to a right to use "for academic purposes", except this terminology is more advantageous because (i) it allows for private companies to advance the general state of the art in the field of ML and AI (e.g. how Google, Facebook and Element AI publish research under traditional open source licenses); (ii) it eliminates the ambiguity around whether university tech transfer offices could commercialise assets developed based on "academic use" licenses; and (iii) it is clear that the restrictions apply both to the original Data and to the other elements which stem from use of that Data. 

The right to use the Data for Internal use would allow the licensee to train the prediction model and trade on proprietary capital (for profit), but not make those predictions available to third parties. 

The Output Commercialization right would allow the licensee to create a SaaS platform that delivers stock predictions (e.g. an API that plugs into customers' trading system, allowing the customer to automatically execute on trade predictions made by the Model) or directly trade third-party capital. However, they would not be able to sell a copy of the Model itself. 

The Model Commercialization right would allow the licensee to do all of the above, as well as sell version of the Model itself. For example, the licensor could sell perpetual licenses to the model so that their end users could use, modify and distribute the Model in any manner they wished. 

\subsection{Additional Restrictions}
As demonstrated above, with a clear understanding of the different uses based on the above terminology, database controllers will be able to be much clearer regarding what is allowed or prohibited using their data, without having to resort to long and often ambiguous license text. In fact, we believe that the relevant licenses could be presented visually in a very clear, concise and uniform way, thus reducing transaction costs when using and making available data. Licensors of data may wish to include additional restrictions. The following may prove useful in such cases:
\begin{itemize}
    \item \textbf{Designated third parties} - We mean to establish use-cases that are agnostic on the actor performing such actions, but in some cases, data licensors may wish to restrict further to specific entities. We note, however, that this may yet reproduce some of the same practical conflicts and tensions we illustrated above (e.g. a university professor may work with data in her or his lab at a University, but may not use the trained models within a commercial context in another side collaboration). Also, the use of Outputs or Trained Models could be limited to specific entities. 
    \item \textbf{Sub-licensability} - a licensor may wish to restrict the right for a licensee to itself make the data itself available (i.e. sub-license it) to other third parties. Implementing such a restriction would also prevent third party contractors from working with the licensed data. 
\item 
\textbf{Attribution / Confidentiality} - a licensor may require that attribution be given in certain contexts in order to publicize or credit the use of certain datasets. Conversely, a licensor may wish to restrict disclosure of the work done with the licensed data, and so would require that use of licensed data be kept confidential. 

\item
\textbf{Ethical Considerations} - other ethical considerations may be added to the rights granted. For example, there could be exclusions related to use in health-related fields, or restrictions against military use.

\end{itemize}

\section{Caveats and Examples of Practical Limitations}
The first caveat in our analysis in this section is that we have chosen to approach dataset licenses agnostically from their underlying data. Indeed, where data consists of artworks, photography or other copyrighted expressions, copyright statutes may provide legal frameworks to consider their use in ML and AI separately from the legal framework governing the collection of data. 

Indeed, in some jurisdictions, authorities have readily and beneficially seized upon reform to clarify use cases relevant to ML and AI, namely through data mining exemptions within copyright statutes. In doing so, some measure of ambiguity with respect to data use in AI is resolved, though, admittedly, initiatives for reform may ultimately be understood to further limit use of data\footnote{For example, initiatives in the European Union to provide for a clear exemption for "text and data mining" lead to the adoption of an exemption that will only apply to "research institutions" in a move that has been qualified as "having a chilling effect on innovation"  and denying "EU startups a level playing field" (Julia Reda, "Text and Data Mining Limited", \url{https://juliareda.eu/eu-copyright-reform/text-and-data-mining/})}. It should also be noted that copyright of the underlying data is not the only issue: the underlying data may be subject to other rights such as privacy rights, image rights and publicity rights. These specific issues can be more difficult to navigate than copyright-related issues for data users (especially since the nature of training implies that the users may not necessarily be analyzing the entirety of a given dataset before combining it with Models, and therefore may not know if there are PII or image issues involved). A striking example is the controversy around the database of photographs drawn from Flickr by IBM, a database that set out to convey more representative data to fight bias in AI\footnote{IBM Research blog, "IBM to release world's largest annotation dataset for studying bias in facial analysis", \url{https://www.ibm.com/blogs/research/2018/06/ai-facial-analytics/}, June 27 2018)}, only to have done so in possible violation of privacy rights\footnote{BBC News, "IBM used Flickr photos for facial-recognition project", \url{https://www.bbc.com/news/technology-47555216}, March 13, 2019.}.

 Additionally, given the importance of these rights, the regulatory frameworks at play tend to be more stringent, hence the use made by users of such datasets may not be exempted under legal exemptions or research-based rationales such as fair use-type exemptions under copyright law which have come to be relied upon by reflex for database creation and/or training. Hence, as a caveat, available datasets may yet suffer from copyright violations, PII violations or other restrictions in their collection, which are not solved by the clear license terms proposed in this article. In sum, a dataset containing ill-acquired data or otherwise breaching privacy rights will not be saved by the terms through which it is offered to the public under license. 

The second caveat is that we do not posit the existence of property rights in data as such. Current legal research is quickly advancing on this front, though we caution here against recognizing rights in data as such, as exposed by \citep{drexl2016data}. 

The third caveat is to note the difference between legislation that is specific to database rights, such as in the EU under the Database Directive (and corresponding implementing national statutes), or under copyright statutes (such as Canada). That said, we believe the analysis and framework of this paper to be removed from this distinction - we aim to describe access rights to databases as such rights are in personam, enforceable contracts, meaning that the existence of specific legal status for databases coexists with contracts to access and make use of the data contained in databases. Given the practical and conceptual limitations set out above, as well as their potential impact on the development of ML and AI, we believe that more adapted and accurate licenses to data are required. This standardization of the taxonomy used to establish rights to use data is more representative of practical realities in the field, and if adopted, may herald much needed clarity. 

\section{Standardization of AI Asset Licenses}
Appendix 2 of this paper presents a comprehensive summary chart of the rights that may be granted by referencing the taxonomy developed in this article. Appendix 3 offers a blank chart that delineates the taxonomy and rights described in this paper. Re-use of this chart is encouraged, as the aim is for it to be used by the AI and ML communities. Doing so would be beneficial in that a common frame of reference could emerge and bring much needed transparency. If users use this "Top Sheet" alongside the license language and the data, others may be able to assess at a glance what rights are granted, thereby decreasing the time to review and understand what can be done with data. 

The authors also make available the licensing language in Appendix 4 under the name "Montreal Data License (MDL)". The aim is to create and drive adoption of this new family of license. Incidentally, the matrix presented at Appendix 3 may be combined with an online tool to automatically generate license terms based on a selection of the different rights that they wish to grant in relation to the different assets with regards to different entities. Such a tool is made available online by the authors and their collaborators at \url{www.montrealdatalicense.com}. 

There are inherent limitations to this framework, but the baseline developed can accommodate more complex use cases, such as licensing of personal information. In fact, doing so under a clearer licensing framework, defined with specific actions that may be taken, may contribute to regulatory compliance. For example, under most privacy statutes and rules, such as the European Union's General Data Protection Regulations, consents from individuals must be informed and specific. By permitting clearer use cases and actions to downstream users, licensors of data may ensure greater clarity as to the scope of authorized actions.

\section{Conclusion}
The emergence of ML and AI is already shaping society, political systems and our economies. The underlying assets driving such changes are largely informational. Access and licensing of data can thus be understood as one of the cornerstone of the development of ML and AI. This is true in an abstract sense, but when combined to the fact that there exists a widening data gap between multinational firms with platform-based business models on one hand, and governments, citizens and other businesses on the other, the need for clarity in data licensing becomes imperative. 

This paper aimed to bring a step forward to bring about this clarity from a legal standpoint by providing a licensing framework anchored in practical realities of ML and AI. The goal is ambitious: providing a common frame of reference to create standards for data licensing to compare with those found in open source software. When combined with recent works such as the two-sided markets conceptualized by Agarwal, \textit{et al.}\citep{agarwal2018marketplace}, one can construct the foundation of markets for data that can help foster transparency, algorithmic fairness and fair market dynamics.

A small step towards this is the creation of the Montreal Data License (MDL) – a modular approach to data licensing in AI and ML that the authors hope will break ground and be adopted by the AI and ML communities. Interested readers may wish to send feedback on this paper, the license language and the License Generator tool by writing via email to \url{info@montrealdatalicense.com}.

\bibliography{main}

\begin{thebibliography}{11}
\providecommand{\natexlab}[1]{#1}
\providecommand{\url}[1]{\texttt{#1}}
\expandafter\ifx\csname urlstyle\endcsname\relax
  \providecommand{\doi}[1]{doi: #1}\else
  \providecommand{\doi}{doi: \begingroup \urlstyle{rm}\Url}\fi

\bibitem[Agarwal et~al.(2018)Agarwal, Dahleh, and
  Sarkar]{agarwal2018marketplace}
Anish Agarwal, Munther Dahleh, and Tuhin Sarkar.
\newblock A marketplace for data: An algorithmic solution.
\newblock \emph{arXiv preprint arXiv:1805.08125}, 2018.

\bibitem[Coates et~al.(2011)Coates, Ng, and Lee]{stl}
Adam Coates, Andrew Ng, and Honglak Lee.
\newblock An analysis of single-layer networks in unsupervised feature
  learning.
\newblock In \emph{Proceedings of the fourteenth international conference on
  artificial intelligence and statistics}, 2011.

\bibitem[Deng et~al.(2009)Deng, Dong, Socher, Li, Li, and
  Fei-Fei]{imagenet_cvpr09}
J.~Deng, W.~Dong, R.~Socher, L.-J. Li, K.~Li, and L.~Fei-Fei.
\newblock {ImageNet: A Large-Scale Hierarchical Image Database}.
\newblock In \emph{CVPR}, 2009.

\bibitem[Drexl et~al.(2016)Drexl, Hilty, Desaunettes, Greiner, Kim, Richter,
  Surblyte, and Wiedemann]{drexl2016data}
Josef Drexl, Reto Hilty, Luc Desaunettes, Franziska Greiner, Daria Kim, Heiko
  Richter, Gintare Surblyte, and Klaus Wiedemann.
\newblock Data ownership and access to data-position statement of the max
  planck institute for innovation and competition of 16 august 2016 on the
  current european debate.
\newblock 2016.

\bibitem[Gebru et~al.(2018)Gebru, Morgenstern, Vecchione, Vaughan, Wallach,
  Daume{\'e}~III, and Crawford]{gebru2018datasheets}
Timnit Gebru, Jamie Morgenstern, Briana Vecchione, Jennifer~Wortman Vaughan,
  Hanna Wallach, Hal Daume{\'e}~III, and Kate Crawford.
\newblock Datasheets for datasets.
\newblock \emph{arXiv preprint arXiv:1803.09010}, 2018.

\bibitem[Krizhevsky \& Hinton(2009)Krizhevsky and Hinton]{cifar}
Alex Krizhevsky and Geoffrey Hinton.
\newblock Learning multiple layers of features from tiny images.
\newblock Technical report, Citeseer, 2009.

\bibitem[Lin et~al.(2014)Lin, Maire, Belongie, Hays, Perona, Ramanan,
  Doll{\'a}r, and Zitnick]{coco}
Tsung-Yi Lin, Michael Maire, Serge Belongie, James Hays, Pietro Perona, Deva
  Ramanan, Piotr Doll{\'a}r, and C~Lawrence Zitnick.
\newblock Microsoft coco: Common objects in context.
\newblock In \emph{ECCV}, 2014.

\bibitem[Maddern et~al.(2017)Maddern, Pascoe, Linegar, and
  Newman]{RobotCarDatasetIJRR}
Will Maddern, Geoff Pascoe, Chris Linegar, and Paul Newman.
\newblock {1 Year, 1000km: The Oxford RobotCar Dataset}.
\newblock \emph{IJRR}, 2017.

\bibitem[Mitchell et~al.(2019)Mitchell, Wu, Zaldivar, Barnes, Vasserman,
  Hutchinson, Spitzer, Raji, and Gebru]{mitchell2019model}
Margaret Mitchell, Simone Wu, Andrew Zaldivar, Parker Barnes, Lucy Vasserman,
  Ben Hutchinson, Elena Spitzer, Inioluwa~Deborah Raji, and Timnit Gebru.
\newblock Model cards for model reporting.
\newblock In \emph{FAT*}, 2019.

\bibitem[Unit(2011)]{Economist}
Economist~Intelligence Unit.
\newblock Economies of scale: How the oil and gas industry cuts costs through
  replication.
\newblock
  \url{https://eiuperspectives.economist.com/sites/default/files/Oil\%20and\%20Gas_\%20Economies\%20of\%20Scale.pdf},
  2011.

\bibitem[Zhang et~al.(2008)Zhang, Sun, and Tang]{cat}
Weiwei Zhang, Jian Sun, and Xiaoou Tang.
\newblock Cat head detection-how to effectively exploit shape and texture
  features.
\newblock In \emph{ECCV}, 2008.

\end{thebibliography}
\bibliographystyle{main}
\newpage
\begin{appendices}
\chapter{Appendix 1}
\begin{table}[h]
\begin{tabular}{ll}
\multicolumn{2}{c}{\textbf{Overview of commonly used datasets}} \\
\\
\hline
\multicolumn{1}{|p{7cm}|}{CIFAR10 and
CIFAR100 datasets~\citep{cifar} \newline (\url{https://www.cs.toronto.edu/~kriz/cifar.html})} & \multicolumn{1}{|p{7cm}|}{To access, view and/or download the Data to view it and evaluate it (evaluation algorithms may be exposed to it, but no Untrained Models).} \\ \hline
\multicolumn{1}{|p{7cm}|}{STL 10 dataset (\citep{stl}) \newline
(\url{https://cs.stanford.edu/~acoates/stl10/})} & \multicolumn{1}{|p{7cm}|}{The site asks that it be cited for all uses of the dataset. Images are pulled from the CIFAR datasets} \\ \hline
\multicolumn{1}{|p{7cm}|} {Cat dataset~\citep{cat}
\newline (\url{https://archive.org/details/CAT_DATASET})} & \multicolumn{1}{|p{7cm}|}{Make all or part of the data available to third parties.} \\ \hline
\multicolumn{1}{|p{7cm}|}{Oxford Robocar dataset~\citep{RobotCarDatasetIJRR} \newline
(\url{http://robotcar-dataset.robots.ox.ac.uk/})} & \multicolumn{1}{|p{7cm}|}{The dataset is available under a non-commercial share alike license that prohibits commercial use. Attribution is required if the dataset is used. They seem to have created the data in the data set themselves \newline \url{http://robotcar-dataset.robots.ox.ac.uk/}} \\ \hline

\multicolumn{1}{|p{7cm}|}{MS COCO labelled image dataset~\citep{coco} \newline
(\url{http://mscoco.org/})} & \multicolumn{1}{|p{7cm}|}{The dataset creators do not have any title to the images, which are pulled from Flickr. They are available under a creative commons license \newline \url{http://mscoco.org/terms_of_use/}} \\ \hline

\multicolumn{1}{|p{7cm}|}{Nexar dataset and challenge \newline
(\url{https://www.getnexar.com/challenge-2/}} & \multicolumn{1}{|p{7cm}|}{The Dataset can only be used for the challenge and can only be used by individuals \newline \url{https://www.getnexar.com/challenge-2/terms-of-use/}} \\ \hline

\multicolumn{1}{|p{7cm}|}{Dogs vs Cats
\newline
\url{https://www.kaggle.com/c/dogs-vs-cats/data}} & \multicolumn{1}{|p{7cm}|}{They may be used for non-commercial research purposes, but they may not be re-published without the express permission of Microsoft Research. The Microsoft research agreement also states that it is for non-commercial research purposes \newline \url{https://github.com/Microsoft/FastRDFStore/blob/master/LICENSE.md}} \\ \hline

\multicolumn{1}{|p{7cm}|}{Imagenet~\citep{imagenet_cvpr09}
\newline
\url{http://image-net.org/download-features}} & \multicolumn{1}{|p{7cm}|}{ImageNet does not own the copyright of the images. For researchers and educators who wish to use the images for non-commercial research and/or educational purposes, we can provide access through our site under certain conditions and terms.} \\ \hline

\multicolumn{1}{|p{7cm}|}{Linguistic Data Consortium (for Non-Members) \newline \url{https://catalog.ldc.upenn.edu/license/ldc-non-members-agreement.pdf}} & \multicolumn{1}{|p{7cm}|}{User agrees to use the LDC Databases received under this Agreement only for non-commercial linguistic education, research and technology development. In the event that User's use of the LDC Databases results in the development of a commercial product, User must join LDC as a For-Profit Member and pay all applicable fees prior to release of said commercial product.} \\ \hline

\end{tabular}
\end{table}
\newpage
\chapter{Appendix 2}
\begin{figure}[h]
\centering
  \caption{Summary of rights granted in conjunction with Models}
  \includegraphics[width=1.1\textwidth]{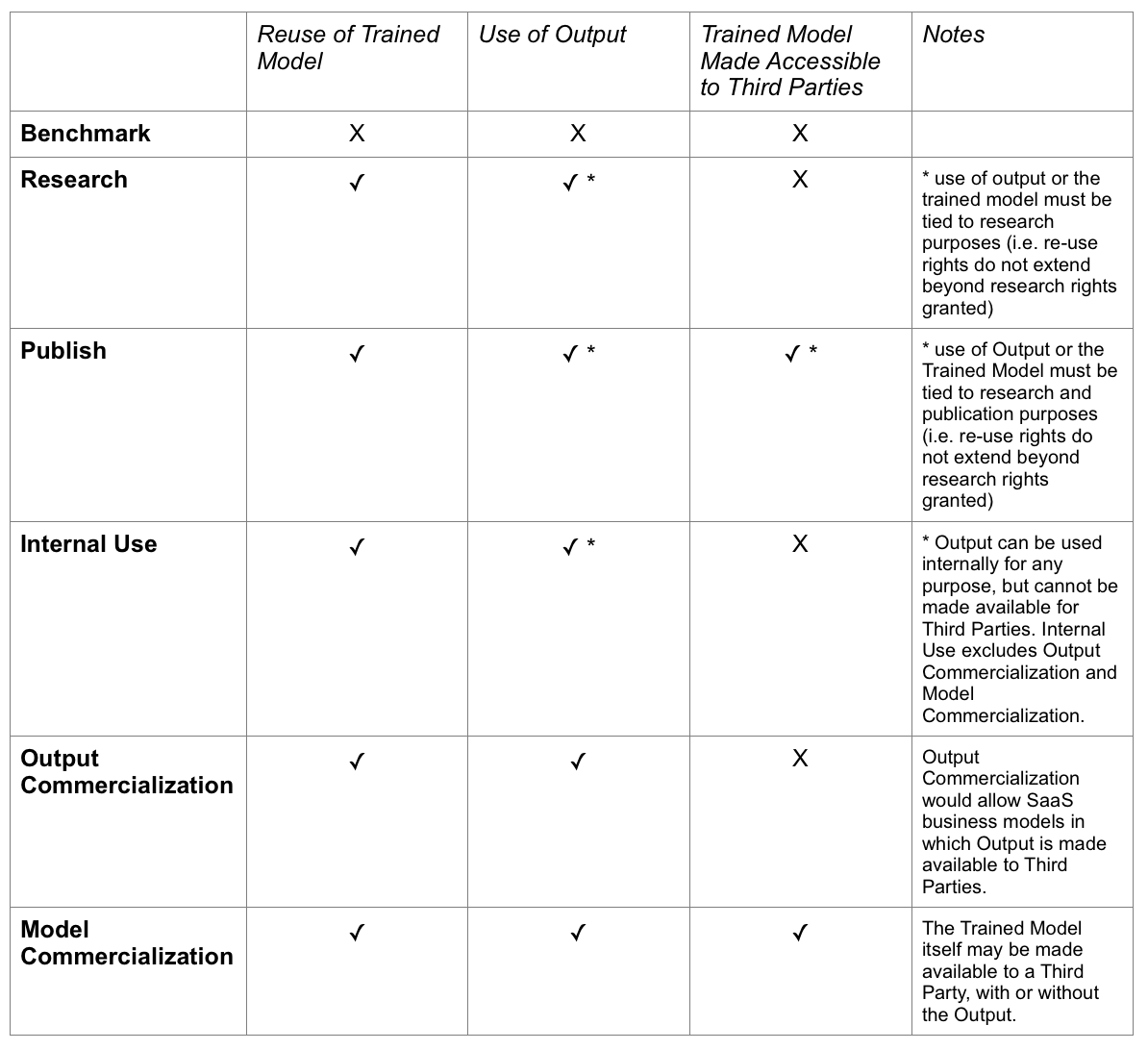}
\end{figure}
\newpage
\chapter{Appendix 3}
\begin{figure}[h]
\centering
  \caption{Top Sheet for Licensed Rights}
  \includegraphics[width=1.1\textwidth]{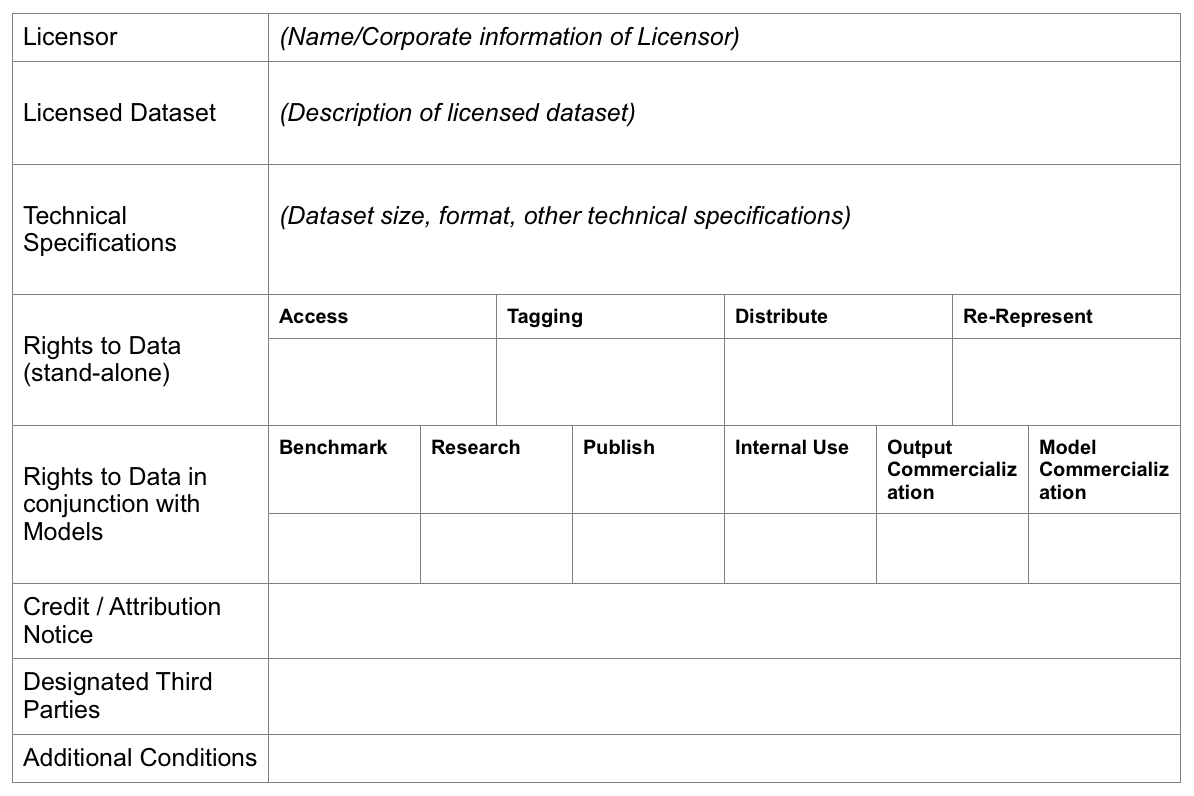}
\end{figure}
\newpage
\chapter{Appendix 4: \textbf{Licence Language}}\label{app4}
\newline
The following licensing language is made available under CC-BY4. Attribution should be made to \textbf{Montreal Data License (MDL)}, or \textbf{License language based on Montreal Data License}. 

The authors are not legal advisors to the individuals and entities making use of these licensing terms. The licensing terms can be combined as needed to match the rights conferred by the licensor. 

The language below assumes that all rights are granted, however each right should be conferred or not based on the user's intent.
\newline
Data License for use in AI and ML:
\newline
This license covers the Data made available by Licensor to you (\textbf{License}) under the following terms. Licensee's use of the data consists acceptance of the terms of this license agreement (\textbf{License}). 

\begin{enumerate}
\item Definitions
\begin{enumerate}[label=(\alph*)]

    \item \textit{Data} means the informational content (individually or as a whole) made available by Licensor.
    \item \textit{Model} means machine-learning or artificial-intelligence based algorithms, or assemblies thereof that, in combination with different techniques, may be used to obtain certain results. Without limitation, such results can be insights on past data patterns, predictions on future trends or more abstract results. 
    \item \textit{Output} means the results of operating a Trained Model as embodied in informational content resulting therefrom. 
    \item \textit{Representation} is a transformation of a piece of data into a different form. Good representations can be used as input to perform useful tasks.
    \item \textit{Labelled Data} means the associated metadata and informational content derived from Data which identify, comment or otherwise derive information from Data, such as tags and labels.
    \item \textit{Licensor} means the individual or entity making the Data available to you.
    \item \textit{Third Parties} means individuals or entities that are not under common control with Licensee.
    \item \textit{Train} means to expose an Untrained Model to the Data in order to adjust the weights, hyperparameters and/or structure thereof.
    \item \textit{Trained Model} means a Model that is exposed to Data such that its weights, parameters and architecture embody insights from the Data. 
    \item \textit{Untrained Model} means Model that is conceived and reduced to practice as to its structure, components and architecture but that has not been trained on Data such that its weights, parameters and architecture do not embody insights from the Data. 

\end{enumerate}

\item General Clauses
\begin{enumerate}[label=(\alph*)]
    \item Unless otherwise agreed in writing by the parties, the data is licensed "as is" and "as available". Licensor excludes all representations, warranties, obligations, and liabilities, whether express or implied, to the maximum extent permitted by law. 
    \item Nothing in this License permits Licensee to make use of Licensor's trademarks, trade names, logos or to otherwise suggest endorsement or misrepresent the relationship between the parties. 
    \item The rights granted under this license are deemed to be non-exclusive, worldwide, perpetual and irrevocable, unless otherwise specified in writing by Licensor. 
    \item Without limiting Licensee's rights available under applicable law, all rights not expressly granted hereunder are hereby reserved by Licensor. The Data and the database under which it is made available remain the property of Licensor (and/or its affiliates or licensors).
    \item This license shall be terminated upon any breach by Licensee of the terms of this License. 
\end{enumerate}

\item Licensed Rights to the Data
\begin{enumerate}[label=(\alph*)]
\item Licensor hereby grants the following rights to Licensee with respect to making use of the Data itself.
\begin{enumerate}[label=(\roman*)]

    \item Access the Data, where "access" means to access, view and/or download the Data to view it and evaluate it (evaluation algorithms may be exposed to it, but no Untrained Models).
    \item Creation of Tagged Data.
    \item Distribute the Data, i.e. to make all or part of the Data available to Third Parties under the same terms as those of this License.
    \item Creation of a Representation of the Data.

\end{enumerate}

\item The rights granted in (a) above exclude the following rights with respect to making use of the Data itself:

\begin{enumerate}[label=(\roman*)]

    \item Any right not granted under (a) should be included in this subsection 
\end{enumerate}
\end{enumerate}
\item Licensed Rights in Conjunction with Models.
\begin{enumerate}[label=(\alph*)]

\item Licensor hereby grants the following rights to Licensee with respect to making use of the Data in conjunction with Models.
\begin{enumerate}[label=(\roman*)]

    \item Benchmark: To access the Data, use the Data as training data to evaluate the efficiency of different Untrained Models, algorithms and structures, but excludes reuse of the Trained Model, except to show the results of the Training. This includes the right to use the dataset to measure performance of a Trained or Untrained Model, without however having the right to carry-over weights, code or architecture or implement any modifications resulting from the Evaluation.
    \item Research: To access the Data, use the Data to create or improve Models, but without the right to use the Output or resulting Trained Model for any purpose other than evaluating the Model Research under the same terms. 
    \item Publish: To make available to Third Parties the Models resulting from Research, provided however that third parties accessing such Trained Models have the right to use them for Research or Publication only.
    \item Internal Use: To access the Data, use the Data to create or improve Models and resulting Output, but without the right to Output Commercialization or Model Commercialization. The Output can be used internally for any purpose, but not made available to Third Parties or for their benefit.
    \item Output Commercialization: To access the Data, use the Data to create or improve Models and resulting Output, with the right to make the Output available to Third Parties or to use it for their benefit, without the right to Model Commercialization.
    \item Model Commercialization: Make a Trained Model itself available to a Third Party, or embodying the Trained Model in a product or service, with or without direct access to the Output for such Third Party.

\end{enumerate}
\item The rights granted in (a) above exclude the following rights with respect to making use of the Data in conjunction with Models:
\begin{enumerate}[label=(\roman*)]

    \item (Any right not granted under (a) should be included in this subsection (b))

\end{enumerate}
\end{enumerate}

\item Attribution and Notice
Attribution and Notice. The origin of the Data and notices included with the Data shall be made available to Third Parties to whom the Data, Output and/Model have been made available. Any distribution of all or part of the Data shall be done under the same terms as those of this License. Licensee shall make commercially reasonable efforts to link to the source of the Data. If so indicated by the Licensor in writing alongside the Data that the use shall be deemed confidential, then Licensee shall not publicly refer to Licensor and/or the source of the Data.
\end{enumerate}
\end{appendices}
\end{document}